\documentclass{article}

\PassOptionsToPackage{numbers, compress}{natbib}
\usepackage[preprint]{neurips_2026}

\usepackage[utf8]{inputenc}
\usepackage[T1]{fontenc}
\usepackage{hyperref}
\usepackage{url}
\usepackage{booktabs}
\usepackage{multirow}
\usepackage{amsfonts}
\usepackage{amsmath}
\usepackage{amssymb}
\usepackage{nicefrac}
\usepackage{microtype}
\usepackage{xcolor}
\usepackage{graphicx}
\usepackage{algorithm}
\usepackage{algorithmic}
\usepackage{tabularx}
\usepackage{array}
\usepackage{enumitem}
\usepackage{wrapfig}
\usepackage{placeins}
\usepackage{float}

\title{SpeakerLLM: A Speaker-Specialized Audio-LLM for Speaker Understanding and Verification Reasoning}

\author{
  KiHyun Nam$^{1}$ \quad
  Jungwoo Heo$^{2}$ \quad
  Siu Bae$^{1}$ \quad
  Ha-Jin Yu$^{2}$ \quad
  Joon Son Chung$^{1\dagger}$ \\[6pt]
  $^{1}$Korea Advanced Institute of Science and Technology (KAIST) \\
  $^{2}$University of Seoul \\[4pt]
  \texttt{nkh.mmai@kaist.ac.kr} \\
  $^{\dagger}$Corresponding author
}

\begin{document}

\maketitle

\begin{abstract}
As audio-first agents become increasingly common in physical AI, conversational robots, and screenless wearables, audio large language models (audio-LLMs) must integrate speaker-specific understanding to support user authorization, personalization, and context-aware interaction. This requires modeling who is speaking, how the voice sounds, and how recording conditions affect speaker cues. Conventional speaker verification systems provide strong scalar scores but little linguistic evidence, while current audio-LLMs and speaker-aware language models have limited ability to organize speaker information beyond binary labels or descriptive profiles. We present SpeakerLLM, a speaker-specialized audio-LLM framework that unifies single-utterance speaker profiling, recording-condition understanding, utterance-pair speaker comparison, and evidence-organized verification reasoning within a natural-language interface.
We construct verification-reasoning targets and a decision-composition policy that separate profile-level evidence from the final same-or-different decision and organize recording condition, profile evidence, and the decision into a structured trace. At its core, SpeakerLLM uses a hierarchical speaker tokenizer designed to capture multiple granularities of speaker evidence. Utterance-level speaker embeddings summarize identity and profile-level cues, whereas frame-level speaker features preserve fine-grained acoustic descriptors. Experiments show that SpeakerLLM-Base improves speaker-profile and recording-condition understanding over general audio-LLMs, while SpeakerLLM-VR preserves strong generated-verdict accuracy and produces decision traces grounded in the supervised verification reasoning schema. We will release the metadata-enriched supervision dataset and target-construction code for reproducibility.
\end{abstract}

\section{Introduction}
\label{sec:intro}

The rapid proliferation of large language models (LLMs) has accelerated a shift toward agentic AI systems that operate through natural interaction channels. In settings such as physical AI, conversational robots, and screenless wearables, voice is often the primary, and sometimes the only, medium for human-agent interaction~\cite{wang2024survey,rakotomalala2021voice,liu2026visionclaw}. In such audio-first environments, speaker-specific understanding is needed to support user authorization, personalization, and context-aware interaction.
Speaker verification (SV) is therefore no longer merely an auxiliary personalization module; it becomes the identity-facing layer of a broader speaker-aware interface.

The shift reframes SV from a backend scoring problem to an audio-language interface problem~\cite{zhang2023speechgpt,chu2023qwen,tang2024salmonn,ghosh2026audio}. Conventional SV systems compare two recordings by producing a similarity score from speaker embeddings~\cite{dehak2011frontend,snyder2018xvector,wan2018ge2e} and thresholding it into a same/different decision. While effective as backend modules, such systems do
not explain whether a rejection is likely caused by transient noise, reverberation, recording mismatch, or genuine speaker-identity mismatch. A speaker-aware audio-LLM should therefore go beyond a bare verdict and organize speaker evidence in natural language so that the decision is auditable to the
user~\cite{miller2019explanation,liao2020questioning}.

Existing research remains fragmented with respect to this goal. Conventional SV systems provide strong verification accuracy but expose only scalar scores without linguistic evidence~\cite{desplanques20_interspeech}. Explainable SV
methods introduce attribute-based concept bottlenecks~\cite{wu2024explainable,koh2020concept} or explainable voice vectors~\cite{lee25e_interspeech}, but remain largely tied to score-based verification. Recent speaker-aware LLMs move toward language-based interaction, but often cast verification as binary
text-label classification over speaker embeddings~\cite{yang2024speaker,yin2026speakerlm}.
Conversely, generative speaker profiling models can describe speaker attributes, but descriptive profiles alone are insufficient for identity verification because multiple speakers can share correct descriptors such as
gender, age, or regional background~\cite{baali2025colmbo,rose2002forensic}.
Thus, a unified framework that connects speaker understanding, utterance-pair comparison, and evidence-organized verification reasoning remains missing.

We present \textbf{SpeakerLLM}, a speaker-specialized audio-LLM framework for natural-language speaker understanding and comparison. SpeakerLLM targets an LLM-interface setting in which speaker-related acoustic cues are read, compared, and expressed in natural language. It is built around two design principles. First, speaker evidence is distributed across representation granularities. Utterance-level speaker embeddings summarize identity and profile-level cues,
whereas frame-level speaker features preserve fine-grained acoustic evidence such as pitch, timbral brightness, and recording condition. We operationalize this representation-granularity observation with a \emph{hierarchical speaker tokenizer} that converts both representations into
continuous speaker tokens for the language model.
Second, speaker cue reading, utterance-pair comparison, and
evidence-organized verification reasoning require different supervision
structures.
We therefore train SpeakerLLM along a two-stage trajectory, yielding
\textbf{SpeakerLLM-Base} for speaker-profile QA, recording-condition QA, and
standard same/different speaker judgment, and \textbf{SpeakerLLM-VR} for
speaker verification reasoning.

A key component of SpeakerLLM-VR is the \textbf{verification reasoning target construction policy}.
Rather than generating free-form rationales~\cite{wei2022chain}, speaker verification reasoning uses a three-block format: \textsc{environment\_status}, \textsc{profile\_compatibility}, and \textsc{decision}.
The target organizes recording condition, profile-level evidence, and the final same/different decision into an evidence-organized decision trace. Crucially, it separates profile-level evidence from the final decision: two
utterances may have similar profiles but come from different speakers, or differ in some profile attributes while belonging to the same speaker. We explicitly include such reversal cases to discourage shortcuts that map profile similarity mechanically to a same-speaker decision. In this work, we operationalize faithfulness in an evidence-grounded sense~\cite{lanham2023measuring,turpin2023language}:
generated evidence blocks are evaluated against the supervised schema used to construct the verification reasoning target.

Our contributions are as follows:
\begin{itemize}
    \item \textbf{Speaker-specialized audio-LLM framework.}
    We introduce SpeakerLLM as a unified natural-language interface for single-utterance speaker profiling, recording-condition understanding, utterance-pair speaker comparison, and evidence-organized verification reasoning.

    \item \textbf{Representation-granularity analysis for speaker evidence.}
    We show that utterance-level speaker embeddings and frame-level speaker features preserve complementary speaker evidence, and operationalize this finding with a hierarchical speaker tokenizer for speaker-specialized audio-LLMs.

    \item \textbf{Evidence-organized reasoning target construction.}
    We design a verification reasoning target and decision composition policy that separate profile-level evidence from the final same/different decision and organize recording condition, profile evidence, and the decision into an auditable trace, including reversal cases that discourage profile-similarity shortcuts.
    We also construct a metadata-enriched supervision dataset and will release the dataset and target-construction code.
\end{itemize}

\begin{figure*}[t!]
\centering
\includegraphics[width=0.95\textwidth]{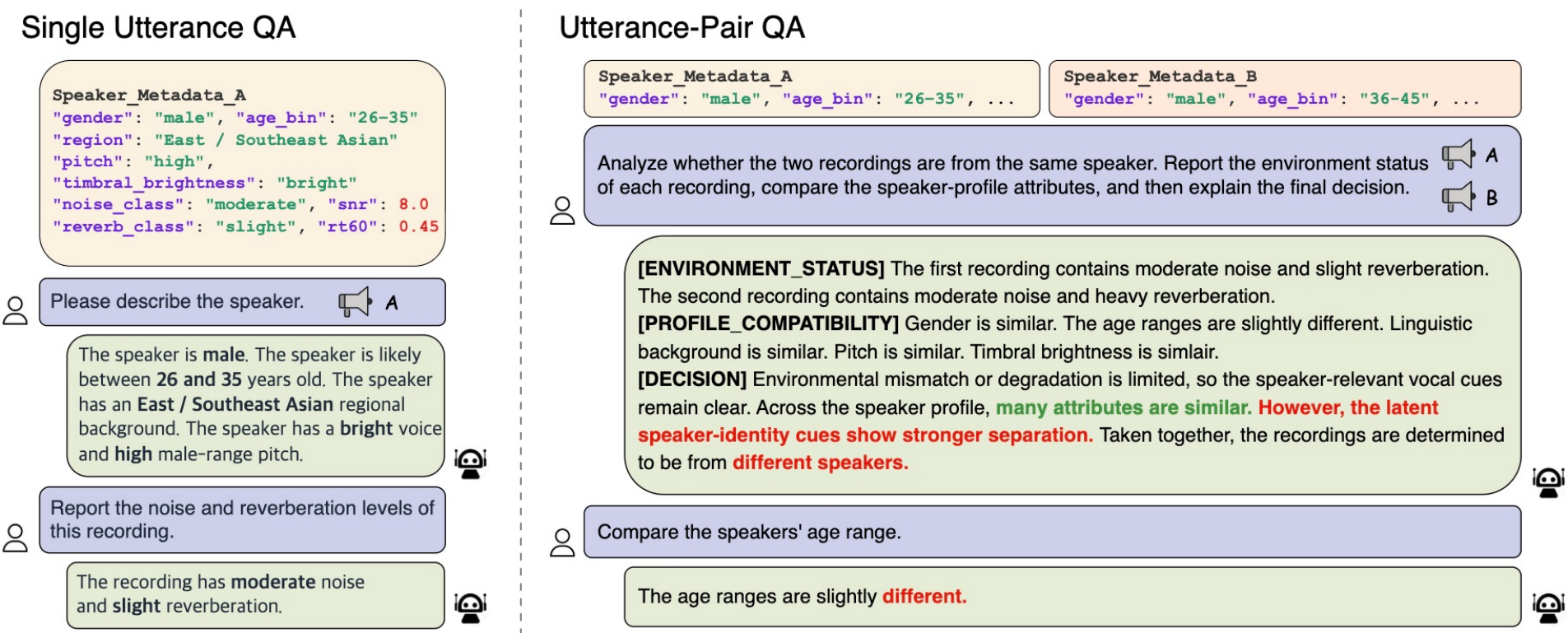}
\vspace{-5pt}
\caption{
QA task inventory for SpeakerLLM training.
Single-utterance tasks read speaker-profile and recording-condition cues;
utterance-pair tasks include standard SV, attribute compatibility QA, and the
three-block verification reasoning target.
}
\label{fig:example_QA_scenarios}
\vspace{-1mm}
\end{figure*}


\section{Speaker Understanding and Verification Reasoning Dataset}
\label{sec:dataset}

We construct a metadata-enriched supervision corpus for speaker-specialized
audio-LLM training.
The corpus connects single-utterance speaker understanding,
recording-condition understanding, utterance-pair speaker comparison, and
evidence-organized verification reasoning in natural language.
Rather than collecting a new corpus from scratch, we combine
VoxCeleb~\cite{nagrani17_interspeech, chung18b_interspeech} and
LibriTTS-R~\cite{koizumi23_interspeech} with public
metadata~\cite{hechmi2021voxceleb, kawamura24_interspeech},
audio-derived descriptors, and online acoustic simulation.
The central design principle is to separate \emph{speaker-profile attributes}
from \emph{environment factors}: the former provide interpretable
speaker-intrinsic evidence, whereas the latter describe recording conditions
that can affect the reliability of that evidence.
The resulting labels define the SpeakerLLM task inventory:
single-utterance QA, utterance-pair QA, and verification reasoning targets.
Table~\ref{tab:data_sources_coverage} summarizes the attribute taxonomy;
details are provided in Appendices~\ref{app:metadata} and~\ref{app:details}.

\begin{table}[t!]
\centering
\scriptsize
\caption{
Attribute taxonomy used in SpeakerLLM.
Speaker-profile attributes are derived from public metadata or extracted
from audio; environment factors are generated online through acoustic
simulation.
Exact binning boundaries are in Appendix~\ref{app:attribute-binning}.
}
\label{tab:data_sources_coverage}
\setlength{\tabcolsep}{6pt}
\begin{tabularx}{\linewidth}{@{}lXl@{}}
\toprule
\textbf{Attribute} & \textbf{Classes} & \textbf{Source} \\
\midrule
Gender
& \{\texttt{male}, \texttt{female}\}
& VoxCeleb, LibriTTS-R \\

Age
& \{\texttt{1--7}, \texttt{8--12}, \texttt{13--17}, \texttt{18--25},
\texttt{26--35}, \texttt{36--45}, \texttt{46--55}, \texttt{56--65},
\texttt{66--75}, \texttt{76+}\}
& VoxCeleb \\

Region
& \{\texttt{North American}, \texttt{European}, \texttt{British / Irish},
\texttt{Latin / Hispanic}, \texttt{Oceanian},
\texttt{East / Southeast Asian}, \texttt{Middle Eastern / North African},
\texttt{African}\}
& VoxCeleb \\

Pitch
& \{\texttt{very low}, \texttt{low}, \texttt{normal},
\texttt{high}, \texttt{very high}\} (gender-conditioned)
& VoxCeleb, LibriTTS-R \\

Timbral brightness
& \{\texttt{muted}, \texttt{mellow}, \texttt{neutral},
\texttt{bright}, \texttt{brilliant}\}
& VoxCeleb, LibriTTS-R \\

Noise
& \{\texttt{clean}, \texttt{mild}, \texttt{moderate},
\texttt{severe}, \texttt{extreme}\}
& MUSAN \\

Reverberation
& \{\texttt{minimal}, \texttt{slight}, \texttt{moderate},
\texttt{heavy}, \texttt{extreme}\}
& SLR28 RIR \\
\bottomrule
\vspace{-5mm}
\end{tabularx}
\end{table}

\subsection{Attribute Supervision}
\label{sec:attribute-supervision}

Speaker-profile attributes are constructed from both metadata-derived profile
labels and audio-derived descriptors.
For VoxCeleb~\cite{nagrani17_interspeech, chung18b_interspeech}, we use VoxCeleb metadata~\cite{hechmi2021voxceleb} to obtain
gender, age, and nationality.
Nationality is mapped to coarser regional/linguistic background classes and
age is converted into coarse bins, providing stable and interpretable cues
instead of optimizing fine-grained demographic prediction from speech.
For LibriTTS-R~\cite{koizumi23_interspeech}, we use only the gender metadata aligned from LibriTTS-P~\cite{kawamura24_interspeech}. We additionally extract pitch (mean F0) and timbral brightness (spectral
centroid) from both VoxCeleb and LibriTTS-R audio; together they form the
voice characteristic target (see Appendix~\ref{app:acoustic-feature-extraction}
for extraction and gender-conditioning details).

Environment factors consist of background noise and reverberation.
Noise is simulated with MUSAN non-speech audio~\cite{snyder2015musan} and
controlled by SNR, while reverberation is simulated with SLR28
RIRs~\cite{slr28RIR} and bucketed by estimated RT60
(Appendix~\ref{app:env-label-computation}).
Although these environment factors do not define speaker identity, they provide
recording-condition evidence that affects the reliability of speaker cues.

\subsection{Training Scenarios}
\label{sec:training-scenarios}

\noindent\textbf{Single-utterance QA.}
Single-utterance QA trains the model to read speaker-profile attributes or
environment factors from one speech recording.
Speaker-side targets include gender, age, regional/linguistic background,
voice characteristic, and full speaker profile.
Environment-side targets include noise, reverberation, and joint acoustic
profile.

\noindent\textbf{Utterance-pair QA.}
Utterance-pair QA uses two recordings as input and supervises relational
speaker evidence.
This category includes attribute compatibility QA, where the model compares a
profile axis across the two utterances, and a standard same/different speaker
judgment.

\noindent\textbf{Verification reasoning target.}
For the evidence-organized reasoning variant, we construct a verification
reasoning target.
The target follows a structured three-block format:
\[
\textsc{environment\_status}
\rightarrow
\textsc{profile\_compatibility}
\rightarrow
\textsc{decision}.
\]
The first block describes the recording conditions of the two utterances.
The second block summarizes whether the available speaker-profile attributes
provide supportive, mixed, or conflicting profile-level evidence.
The final \textsc{decision} block produces the final verification verdict.
Crucially, the decision block is not a direct restatement of the profile-level
evidence: two utterances may have similar profiles but come from different
speakers, or differ in some profile attributes while belonging to the same
speaker.
The verification reasoning target is therefore constructed to separate profile-level evidence from the final same/different decision, allowing the target to express both aligned and reversal cases in a linguistically coherent
form.
Exact target templates, profile-support scoring, and the verification-reasoning
target construction policy are provided in Appendix~\ref{app:details}.
Because full speaker-profile descriptions and speaker verification reasoning require all profile
attributes, these targets are instantiated only on VoxCeleb utterances, where
age and region metadata are available; the \textsc{environment\_status} block
is populated by applying the same online noise and reverberation simulation
used for standalone environment QA on LibriTTS-R.

\begin{figure}[t]
    \centering
    \includegraphics[width=0.85\linewidth]{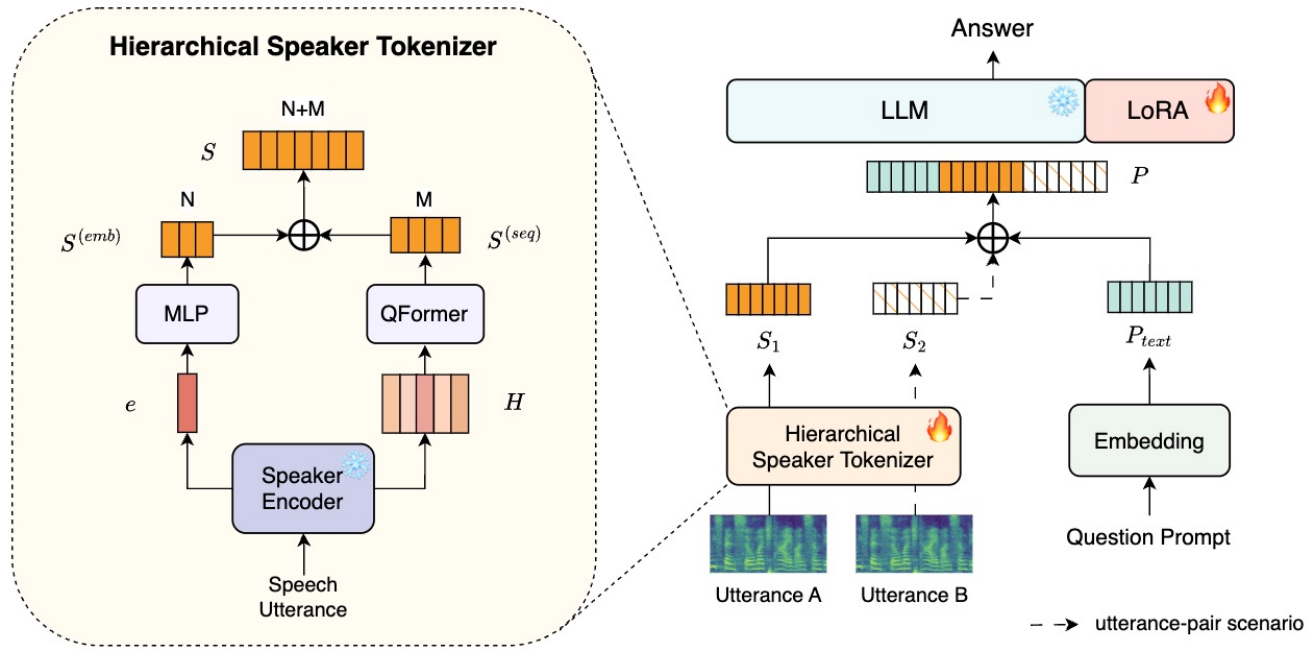}
    \caption{
Overview of SpeakerLLM.
A frozen speaker encoder extracts a speaker embedding $\mathbf{e}$ and
frame-level features $\mathbf{H}$.
The hierarchical speaker tokenizer converts them into embedding-level and
sequence-level speaker tokens, which are inserted at fixed prompt slots for
LLM conditioning.
}
\label{fig:architecture}
\end{figure}
\vspace{-5pt}

\FloatBarrier

\section{Approach}
\label{sec:approach}

SpeakerLLM is a speaker-specialized audio-LLM that reads speaker-related cues
from one or two utterances and maps them to natural-language speaker profiles,
recording-condition descriptions, utterance-pair comparisons, and evidence-organized verification reasoning. As shown in Figure~\ref{fig:architecture}, the model consists of a frozen speaker encoder, a speaker tokenizer, and a language model. The speaker encoder extracts speaker-discriminative representations from the input waveform, the speaker tokenizer converts them into continuous speaker tokens, and the language model consumes these speaker tokens together with textual prompts to generate task-specific answers.

SpeakerLLM is built around two design principles motivated by empirical
observations about speaker representations and training dynamics.
First, speaker-profile attributes and audio-derived descriptors are not
uniformly represented at a single granularity.
Utterance-level speaker embeddings provide stable summaries of speaker identity
and profile-level cues, whereas fine-grained acoustic evidence such as pitch,
timbral brightness, and recording condition is better retained in frame-level
representations.
Second, speaker cue reading, utterance-pair comparison, and evidence-organized
verification reasoning share the same speaker-token conditioning mechanism but
require different supervision structures and output formats.
We therefore train SpeakerLLM along a sequential trajectory that first builds
speaker understanding and then further tunes the same speaker-token
conditioning mechanism for verification reasoning.

The trajectory yields two model variants.
\textbf{SpeakerLLM-Base} is obtained after Stage~1 and is used for
speaker-profile QA, recording-condition QA, and standard same/different
speaker judgments.
\textbf{SpeakerLLM-VR} is obtained by further tuning SpeakerLLM-Base with
verification reasoning targets and is used for verification reasoning.
Both variants share the same frozen speaker encoder, speaker tokenizer,
language-model backbone, and prompt conditioning mechanism.

\subsection{Hierarchical Speaker Tokenizer}
\label{sec:tokenizer}

We use the term \emph{speaker tokenizer} to denote the waveform-to-token
module that converts an input utterance into continuous speaker tokens readable
by the language model.
It consists of a frozen speaker encoder and a trainable speaker-token adapter.
The speaker encoder extracts speaker representations from
the input waveform, and the adapter maps the encoder outputs into
a sequence of speaker tokens in the LLM hidden space.
Thus, the adapter is the learnable mapping component inside the speaker
tokenizer, while the speaker tokenizer denotes the full speaker-specialized
conditioning module.

This distinction is important for speaker-specialized audio-LLMs.
A generic audio-to-LLM connector can be viewed largely as a projection from an
encoder output to the LLM hidden dimension.
For speaker understanding and SV Reasoning, however, dimensional alignment
alone is insufficient.
The final speaker embedding of a speaker verification encoder provides a
compact utterance-level identity summary, but temporal pooling can suppress
fine-grained acoustic evidence needed for explanation, such as pitch, timbral
brightness, and recording condition.
Conversely, frame-level speaker features retain temporal and acoustic detail,
but may lack a compact identity-level summary by themselves.
The observation motivates our \emph{representation-granularity hypothesis}: speaker
evidence useful for an audio-LLM is distributed across multiple representation
levels.

To retain these complementary sources of evidence, SpeakerLLM uses a
hierarchical speaker tokenizer.
Given an input utterance \(\mathbf{x}\in\mathbb{R}^{T}\), the frozen speaker
encoder produces a speaker embedding
\(\mathbf{e}\in\mathbb{R}^{d_e}\) and frame-level speaker features
\(\mathbf{H}\in\mathbb{R}^{L\times d_h}\) before temporal pooling.
The speaker-token adapter processes these two representations with separate
branches.
The embedding-level branch maps \(\mathbf{e}\) to \(N\) speaker tokens with an
MLP, while the sequence-level branch summarizes \(\mathbf{H}\) into \(M\)
speaker tokens with a Q-Former~\cite{li2023blip}:
\begin{align}
\mathbf{S}^{(\mathrm{emb})}
&=
\mathrm{MLP}(\mathbf{e})
\in \mathbb{R}^{N\times d},
\qquad
\mathbf{S}^{(\mathrm{seq})}
=
\mathrm{QFormer}(\mathbf{H})
\in \mathbb{R}^{M\times d}, \\
\mathbf{S}
&=
[\,\mathbf{S}^{(\mathrm{emb})};\mathbf{S}^{(\mathrm{seq})}\,]
\in \mathbb{R}^{(N+M)\times d}.
\end{align}
Here, \(\mathbf{S}\) is the continuous speaker-token sequence inserted into
the LLM prompt.
The hierarchical construction avoids compressing all speaker evidence into a
single embedding and instead retains both embedding-level and sequence-level
speaker evidence in the language-model context.
A token-count ablation verifies that this gain is not attributable to using more speaker tokens alone (Appendix~\ref{app:tokenizer-token-count}).

\subsection{Two-Stage Training for SpeakerLLM Variants}
\label{sec:training-variants}

SpeakerLLM is trained along a single sequential trajectory that yields
SpeakerLLM-Base and SpeakerLLM-VR.
The two-stage procedure is not merely a training schedule; it stabilizes and then
extends the same speaker-token conditioning mechanism for different
interaction modes.

\noindent\textbf{Stage~1: Speaker Understanding.}
Stage~1 trains the model to read speaker-profile attributes, environment
factors, and simple utterance-pair relations.
It includes single-utterance QA and simple utterance-pair QA such as standard
same/different speaker judgment, noise comparison, and reverberation
comparison.

At the beginning of Stage~1, the speaker encoder and language model are kept
frozen, and only the speaker tokenizer is trained with short-form targets.
The warm-up phase focuses on predicting the answer value itself
instead of generating full sentences, thereby aligning audio-derived speaker
representations with the language model input space.
The model is then trained with sentence-form targets while the speaker tokenizer
remains trainable and the language model is adapted with LoRA~\cite{hu2022lora}.
Sentence adaptation then transfers class-style cue reading into
natural-language speaker and environment descriptions.
The resulting model is used as SpeakerLLM-Base.

\noindent\textbf{Stage~2: Verification-Reasoning Tuning.}
Stage~2 further tunes SpeakerLLM-Base for evidence-organized verification
reasoning.
Its key distinction is not the use of paired inputs alone, since Stage~1
already includes simple utterance-pair QA through standard same/different
judgment and environment comparison.
Instead, Stage~2 trains the model to compose recording condition,
profile-level evidence, and the final same/different decision into an
evidence-organized decision trace.

To this end, Stage~2 focuses on verification reasoning targets and attribute
compatibility QA.
The verification reasoning target follows the three-block format
(\textsc{environment\_status}, \textsc{profile\_compatibility},
\textsc{decision}) defined in Section~\ref{sec:training-scenarios}.
Attribute compatibility QA asks the model to compare profile axes such as
gender, age, voice characteristic, regional/linguistic background, and the
overall speaker profile.
A small fraction of sentence-form understanding tasks is also replayed to
maintain the grounded cue-reading ability learned in Stage~1.
Stage~2 therefore serves as a tuning stage that adds verification reasoning
ability on top of SpeakerLLM-Base, without replacing the
speaker-understanding variant.

\subsection{Prompt Conditioning and Objective}
\label{sec:objective}

For each task, the textual prompt is converted into a token-embedding sequence
\(\mathbf{P}_{\text{text}}\) in the language-model hidden space.
For the active utterance set \(\{\mathbf{x}_i\}_{i=1}^{m}\), where
\(m\in\{1,2\}\), the speaker tokenizer produces one speaker-token block per
utterance,
\[
\mathbf{S}_i \in \mathbb{R}^{K_{\mathrm{spk}}\times d},
\qquad
K_{\mathrm{spk}}=N+M.
\]
These speaker-token blocks are inserted into the designated prompt positions,
yielding the prompt-side embedding sequence
\[
\mathbf{P}
=
\mathrm{Insert}
\left(
\mathbf{P}_{\text{text}},
\{\mathbf{S}_i\}_{i=1}^{m}
\right).
\]
The conditioning mechanism is shared across single-utterance QA and
utterance-pair QA: one speaker-token block is inserted for single-utterance
tasks, and two blocks are inserted for utterance-pair tasks.

Given a target sequence \(\mathbf{y}=(y_1,\dots,y_L)\), SpeakerLLM is trained
with the standard autoregressive next-token cross-entropy objective:
\[
\mathcal{L}_{\mathrm{CE}}
=
-\sum_{t=1}^{L}
\log p_{\theta}(y_t \mid y_{<t}, \mathbf{P}).
\]
During speaker-tokenizer warm-up, only the speaker tokenizer is updated.
During sentence adaptation and Verification-Reasoning Tuning, the speaker
tokenizer and LoRA parameters are trained jointly.
The speaker encoder remains frozen throughout.
\FloatBarrier


\section{Experiments}
\label{sec:experiments}

\subsection{Experimental Settings}
\label{sec:exp-settings}

\paragraph{Data.}
We train SpeakerLLM on the VoxCeleb1 development set~\cite{nagrani17_interspeech}
and the LibriTTS-R clean-360h training list~\cite{koizumi23_interspeech},
containing 265k utterances and 530.8 hours of measured audio.
We use this controlled base-scale setting for all main comparisons and
ablations, so that differences across model variants reflect the speaker
tokenizer, training trajectory, and verification reasoning target design,
not changes in data scale.
We center speaker-profile QA and standard same/different speaker judgment
evaluation on VoxCeleb1-O because the VoxCeleb domain provides both
speaker-profile metadata/enrichment and established same/different trials,
allowing profile evidence and speaker judgments to be evaluated in the same
utterance domain.
Environment QA is evaluated on LibriTTS-R test-clean with controlled noise and
reverberation labels.
During training, each audio input is randomly cropped to a duration between
3 and 15 seconds.
At evaluation, we use a fixed 15-second window when the utterance is longer
than 15 seconds, and otherwise use the full utterance.

\paragraph{Model.}
SpeakerLLM uses a frozen ReDimNet-B3~\cite{yakovlev24_interspeech} speaker
encoder and Qwen2.5-1.5B-Instruct~\cite{qwen25report} as the language-model
backbone.
The hierarchical speaker tokenizer maps the final speaker embedding to
16 embedding-level speaker tokens via an MLP and the frame-level speaker
features to 32 sequence-level speaker tokens via a
Q-Former~\cite{li2023blip}, yielding 48 speaker tokens per utterance.
The speaker encoder remains frozen throughout training, while the language
model is adapted with LoRA~\cite{hu2022lora} after the initial speaker-tokenizer warm-up.
LoRA uses rank 16, alpha 32, and dropout 0.05.

\paragraph{Training.}
Training follows the two-stage trajectory in Section~\ref{sec:training-variants}.
Stage~1 first trains only the speaker tokenizer with short-form targets and
then adapts the language model with LoRA using sentence-form speaker
understanding targets, producing \textbf{SpeakerLLM-Base}.
Stage~2 further tunes SpeakerLLM-Base with verification reasoning targets,
attribute compatibility QA, and a small replay of sentence-form understanding
tasks, producing \textbf{SpeakerLLM-VR}.
The three training phases run for 126k, 83k, and 221k steps with AdamW~\cite{loshchilov2018decoupled},
stage-specific peak learning rates of $1{\times}10^{-4}$,
$4{\times}10^{-6}$, and $6{\times}10^{-6}$, bf16 mixed precision, and
FlashAttention-2~\cite{dao2024flashattention}.

\vspace{-2mm}
\paragraph{Evaluation and baselines.}
We report generated-answer accuracy for all LLM-interface tasks.
Generated answers are mapped to task labels with deterministic parsers, and
parsing failures are counted as incorrect.
For utterance-pair speaker judgment, we distinguish two evaluation modes.
\textbf{SV} denotes the standard same/different speaker judgment mode, where
the model generates a short verdict.
\textbf{SV-R} denotes the evidence-organized reasoning mode, where the model
generates a full verification reasoning output; SV-R accuracy is computed from
the same/different verdict parsed from the generated \textsc{decision} block.
SV is evaluated as generated-verdict accuracy rather than score-based EER or
minDCF, because our LLM-interface protocol evaluates textual verdicts and
evidence traces rather than calibrated continuous trial scores.
Metadata is used only for label construction and offline analysis, and is not
provided to any model at inference time.
We compare against general audio-LLMs, including
Qwen2.5-Omni-7B~\cite{chu2023qwen}, Qwen3.0-Omni-30B~\cite{xu2025qwen3},
and Audio Flamingo3~\cite{ghosh2026audio}, using prompted evaluation with
explicit answer options.
We also list SA-TinyLLaMA~\cite{thebaud2026speaker} and
CoLMbo~\cite{baali2025colmbo} for scope comparison; unsupported tasks or
metrics that are not comparable under our generated-verdict protocol are
marked as ``--''.

Full experimental details are provided in Appendix~\ref{app:exp-settings}.

\begin{table}[t!]
\centering
\footnotesize
\caption{
Main results on VoxCeleb1-O (SV and speaker-profile QA) and LibriTTS-R
test-clean (environment QA).
All values are accuracy (\%\,$\uparrow$); bold indicates the best result in
each column.
``--'' denotes unsupported tasks or metrics not comparable under our prompted
generated-answer accuracy protocol.
}
\label{tab:main_results}
\setlength{\tabcolsep}{5pt}
\begin{tabular}{@{}ll c ccccc cc@{}}
\toprule
& &
& \multicolumn{5}{c}{\textbf{Speaker Profile}}
& \multicolumn{2}{c}{\textbf{Environment}} \\
\cmidrule(lr){4-8} \cmidrule(l){9-10}
\textbf{Model}
& \textbf{LLM Backbone}
& SV
& Gender
& Age
& Region
& Pitch
& Bright.
& Noise
& Reverb \\
\midrule
\multicolumn{10}{@{}l}{\textit{General Audio-LLMs}} \\[2pt]
Qwen2.5-Omni-7B~\cite{chu2023qwen}
  & Qwen2.5-7B~\cite{qwen25report}
  & 65.2 & 99.8 & 17.5 & 76.0 & 22.7 & 25.0 & 20.4 & 20.2 \\
Qwen3.0-Omni-30B~\cite{xu2025qwen3}
  & Qwen3-30B-A3B
  & 54.0 & 99.1 & 20.3 & 75.7 & 23.8 & 32.1 & 32.7 & 20.9 \\
Audio Flamingo3~\cite{ghosh2026audio}
  &  Qwen2.5-7B~\cite{qwen25report}
  & 54.5 & 99.9 & 23.2 & 59.8 & 22.1 & 38.3 & 21.2 & 21.0 \\
\midrule
\multicolumn{10}{@{}l}{\textit{Speaker-Specialized LLMs}} \\[2pt]
SA-TinyLLaMA~\cite{thebaud2026speaker}
  & TinyLLaMA-1.1B~\cite{zhang2024tinyllama}
  & -- & -- & -- & -- & -- & -- & -- & -- \\
CoLMbo~\cite{baali2025colmbo}
  & GPT2-base~\cite{radford2019language}
  & -- & 78.6 & 22.7 & 45.5 & -- & -- & -- & -- \\
\midrule
\multicolumn{10}{@{}l}{\textit{Ours}} \\[2pt]
SpeakerLLM-Base
  & Qwen2.5-1.5B~\cite{qwen25report}
  & \textbf{96.1} & \textbf{99.9} & \textbf{39.8} & \textbf{83.1} & \textbf{72.4} & \textbf{54.2} & \textbf{52.7} & \textbf{51.7} \\
\bottomrule
\end{tabular}
\end{table}
\vspace{-5pt}

\subsection{Results}
\vspace{-1mm}
\label{sec:results}

\noindent\textbf{Speaker understanding performance.}
Table~\ref{tab:main_results} summarizes the main LLM-interface benchmark.
We test whether a speaker-specialized audio-LLM can jointly support
speaker-profile, recording-condition, and same/different speaker judgment
within a single natural-language interface.
General audio-LLMs perform well on attributes that are relatively easy to
infer from speech, most notably gender.
However, their performance drops substantially on tasks that require
speaker-specific acoustic evidence or recording-condition awareness, such as
age, pitch, timbral brightness, noise, and reverberation.
This suggests that general audio-LLMs can capture speech content and some
surface paralinguistic cues, but are not designed to retain and interpret
fine-grained speaker-identity evidence or voice-quality differences between
speakers.

Among speaker-specialized LLMs, CoLMbo supports only limited profiling
without utterance-pair judgment, and SA-TinyLLaMA reports score-based EER not
directly comparable to our generated-verdict protocol.
\begin{table}[t!]
\centering
\scriptsize
\caption{
Ablation on speaker-token adapter design.
All variants use the same frozen encoder and LLM, trained under
Stage~1 speaker-tokenizer-only training.
Token-count analysis is in Appendix~\ref{app:tokenizer-token-count}.
}
\label{tab:speaker_tokenizer}
\setlength{\tabcolsep}{5pt}
\begin{tabular}{@{}l cc c ccccc cc@{}}
\toprule
& \multicolumn{2}{c}{\textbf{Speaker Repr.}}
&
& \multicolumn{5}{c}{\textbf{Speaker Profile}}
& \multicolumn{2}{c}{\textbf{Environment}} \\
\cmidrule(lr){2-3} \cmidrule(lr){5-9} \cmidrule(l){10-11}
\textbf{Adapter}
& Embed.
& Frames
& SV
& Gender
& Age
& Region
& Pitch
& Bright.
& Noise
& Reverb \\
\midrule
Linear
  & \checkmark &            & 49.1 & 77.3 & 16.3 & 49.5 & 17.3 & 18.1 & 23.9 & 22.8 \\
MLP
  & \checkmark &            & 86.5 & 99.1 & 30.2 & 78.1 & 57.4 & 41.3 & 32.6 & 32.5 \\
Q-Former
  &            & \checkmark & 84.7 & 99.3 & 32.2 & 74.4 & 70.0 & 51.0 & 48.9 & 45.6 \\
\midrule
Ours (MLP\,+\,Q-Former)
  & \checkmark & \checkmark & \textbf{95.6} & \textbf{99.7} & \textbf{39.5} & \textbf{79.8} & \textbf{72.3} & \textbf{53.1} & \textbf{47.7} & \textbf{50.0} \\
\bottomrule
\end{tabular}
\end{table}

SpeakerLLM-Base achieves the best performance on most speaker-profile and
environment attributes, while also obtaining 96.1\% generated-verdict accuracy
for standard same/different speaker judgment.
The gains are especially large for pitch and timbral brightness, indicating
that targeted speaker supervision is important for voice-characteristic
understanding.
SpeakerLLM-Base also exceeds 50\% accuracy on both noise and reverberation,
which are five-way classification tasks with a chance level of roughly 20\%.
This suggests that the model learns recording-condition cues beyond speech-content understanding.

\noindent\textbf{Speaker tokenizer analysis.}
Table~\ref{tab:speaker_tokenizer} compares the adapter design inside the
speaker tokenizer.
Linear and MLP consume only the final speaker embedding, Q-Former consumes
only frame-level speaker features, and the hierarchical speaker tokenizer
combines both representations.
All variants use the same frozen speaker encoder and LLM backbone and are
evaluated under Stage~1 speaker-tokenizer-only training.

The Linear adapter provides a simple projection lower bound and performs poorly
across tasks, indicating that dimension matching alone is insufficient for
speaker-specialized audio-LLM conditioning.
The MLP adapter retains useful global speaker information from the pooled
speaker embedding, but is limited on attributes requiring temporal or acoustic
detail, such as pitch and timbral brightness.
Conversely, the Q-Former adapter improves these acoustic attributes by
attending to frame-level speaker features, but trails in standard
same/different speaker judgment.
The hierarchical tokenizer achieves the best balance across speaker-profile,
recording-condition, and same/different judgment tasks, supporting our
representation-granularity hypothesis: speaker identity, profile-level cues,
and acoustic voice descriptors benefit from different levels of speaker
representation.
Appendix~\ref{app:tokenizer-token-count} further shows that this gain persists
when MLP-only and Q-Former-only variants are given the same total number of
speaker tokens, confirming that the benefit comes from combining complementary
representation levels, not from token count alone.

\begin{wraptable}{r}{0.52\textwidth}
\vspace{-14pt}
\centering
\footnotesize
\caption{
SV-R evidence grounding on VoxCeleb1-O.
Format Valid measures three-block schema validity; Attr.-level and
Prof.-support measure profile-evidence agreement at clause and summary levels.
All values are accuracy (\%\,$\uparrow$).
}
\label{tab:structured_sv}
\setlength{\tabcolsep}{4pt}
\begin{tabular}{@{}l c cc@{}}
\toprule
& \textbf{Schema}
& \multicolumn{2}{c}{\textbf{Profile Evidence Grounding}} \\
\cmidrule(lr){2-2} \cmidrule(l){3-4}
& Fmt.\ Valid & Attr.-level & Prof.-support \\
\midrule
SpeakerLLM-VR & \textbf{100.0} & \textbf{72.7} & \textbf{63.6} \\
              &                &               & \textit{(maj.\ 52.9)} \\
\bottomrule
\end{tabular}
\vspace{-10pt}
\end{wraptable}

\noindent\textbf{Verification reasoning quality.}
We next evaluate SpeakerLLM-VR, the model variant obtained by further tuning SpeakerLLM-Base with verification reasoning targets.
SpeakerLLM-VR generates an evidence-organized decision trace containing environment status, profile-level evidence, and the final same/different decision.
We evaluate this output from two complementary perspectives: whether the
generated trace is grounded in the supervised evidence schema, and how the
structured trace affects the final generated verdict across trial subsets.

Table~\ref{tab:structured_sv} evaluates the grounding quality of the generated
SV-R trace. SpeakerLLM-VR produces valid three-block outputs in all trials
(\emph{Format Valid} = 100.0\%), indicating reliable adherence to the prescribed evidence schema.
At the clause level, individual profile-comparison statements reach 72.7\%
attribute-level grounding against the supervised evidence labels.
At the summary level, profile-support grounding compares the generated
profile evidence with the target three-way profile-support state
(\textsc{supportive}, \textsc{mixed}, \textsc{conflicting}).
SpeakerLLM-VR reaches 63.6\% profile-support grounding, exceeding the 52.9\%
majority baseline.
The summary-level score is lower because it aggregates multiple clause-level
comparisons into a single three-way state, making it a stricter metric.
Together, full format validity and grounding at both clause and summary levels
support evidence-grounded faithfulness to the supervised verification
reasoning schema.

Table~\ref{tab:svr_diagnostics} provides a subset diagnostic of the final generated verdict.
SV-R maintains the overall generated-verdict accuracy while adding an evidence-organized decision trace, improving the overall score by $+0.33$ percentage points.
The gain is larger on different-speaker trials ($+0.96$ points), and is most
pronounced on supportive-profile different trials ($+1.47$ points), where the
available profile evidence is compatible but the speakers are different.
This subset corresponds to profile-deceptive negative trials, where surface
profile similarity can otherwise encourage a short verdict to predict ``same''.
Small decreases on same-speaker and conflicting-profile same-speaker subsets
reflect near-ceiling fluctuations, as these subsets already exceed 97--99\%
under the short verdict mode.
Thus, SV-R is not intended as a uniformly stronger classifier than direct
verdict generation.
Rather, it maintains final verdict quality overall while exposing an
evidence-organized trace, with the largest gains appearing where
profile-similarity shortcuts are most harmful.
Representative generated traces are provided in
Appendix~\ref{app:qual-examples}.
\begin{table}[t]
\centering
\scriptsize
\caption{
Subset diagnostic for generated same/different verdicts on VoxCeleb1-O.
SV uses the short verdict mode, whereas SV-R parses the verdict from
the generated \textsc{decision} block.
Aligned, mixed, and reversal splits are defined by the relation between
profile-support level and the ground-truth label.
Supp$\rightarrow$diff and Confl$\rightarrow$same denote the two reversal subsets.
$\Delta$ denotes SV-R minus standard SV accuracy.
All values are accuracy (\%\,$\uparrow$).
}
\label{tab:svr_diagnostics}
\setlength{\tabcolsep}{3.5pt}
\begin{tabular}{@{}l cc ccc cc c@{}}
\toprule
& \multicolumn{2}{c}{\textbf{GT-label split}}
& \multicolumn{3}{c}{\textbf{Profile-support split}}
& \multicolumn{2}{c}{\textbf{Hardest reversal}}
& \\
\cmidrule(lr){2-3} \cmidrule(lr){4-6} \cmidrule(lr){7-8}
\textbf{SpeakerLLM-VR}
& Diff. & Same
& Aligned & Mixed & Reversal
& Supp${\to}$diff & Confl${\to}$same
& Overall \\
\midrule
SV
& 94.25 & \textbf{99.32}
& 97.31 & 95.22 & \textbf{97.15}
& 78.53 & \textbf{99.11}
& 96.79 \\
SV-R
& \textbf{95.20} & 99.03
& \textbf{97.87} & \textbf{95.32} & 96.81
& \textbf{80.00} & 98.58
& \textbf{97.12} \\
\midrule
$\Delta$
& \textbf{+0.96} & $-$0.30
& \textbf{+0.56} & \textbf{+0.10} & $-$0.34
& \textbf{+1.47} & $-$0.53
& \textbf{+0.33} \\
\bottomrule
\end{tabular}
\end{table}

\begin{wraptable}{r}{0.46\textwidth}
\vspace{-18pt}
\centering
\footnotesize
\caption{
Training ablation.
(a) Tokenizer warm-up for Stage~1.
(b) Task trajectory for VR Tuning.
Profile and Environ.\ denote mean accuracies.
}
\label{tab:curriculum_ablation}
\setlength{\tabcolsep}{5pt}
\begin{tabular}{@{}lccc@{}}
\toprule
& \multicolumn{3}{c}{\textbf{(a) Tokenizer warm-up}} \\
\cmidrule(l){2-4}
\textbf{Setting} & SV & Profile & Environ. \\
\midrule
w/o warm-up     & 91.20 & 67.16 & 40.46 \\
SpeakerLLM-Base & \textbf{96.05} & \textbf{72.90} & \textbf{52.21} \\
\midrule
& \multicolumn{3}{c}{\textbf{(b) Task trajectory}} \\
\cmidrule(l){2-4}
\textbf{Setting} & \multicolumn{2}{c}{SV} & SV-R \\
\midrule
w/o trajectory  & \multicolumn{2}{c}{\textbf{97.09}} & 91.71 \\
SpeakerLLM-VR   & \multicolumn{2}{c}{96.79} & \textbf{97.12} \\
\bottomrule
\end{tabular}
\vspace{-10pt}
\end{wraptable}

\noindent\textbf{Training ablation.}
Table~\ref{tab:curriculum_ablation}(a) shows that removing the
speaker-tokenizer warm-up substantially degrades the Stage~1 variant across
standard SV, speaker-profile QA, and environment QA, indicating that the
speaker tokenizer benefits from an initial alignment phase before
language-model adaptation.
A per-attribute breakdown is provided in
Appendix~\ref{app:tokenizer-warmup-ablation}.

Table~\ref{tab:curriculum_ablation}(b) shows that mixing Stage~1 and Stage~2
tasks immediately after tokenizer warm-up achieves marginally higher standard SV accuracy
but sharply degrades SV-R accuracy.
Thus, high standard same/different accuracy alone does not imply that the model
has learned the SV-R interface; SpeakerLLM-VR achieves the best SV-R accuracy
when further tuned after the speaker-understanding variant.

\FloatBarrier

\vspace{-3mm}
\section{Conclusion}
\label{sec:conclusion}
\vspace{-3mm}

We presented SpeakerLLM, a speaker-specialized audio-LLM framework that
unifies speaker-profile understanding, recording-condition understanding,
utterance-pair speaker comparison, and evidence-organized verification
reasoning in a natural-language interface.
Our results show that multi-granularity speaker conditioning improves speaker
understanding, and that verification reasoning targets with profile-support
computation and reversal-aware decision composition produce valid,
schema-grounded decision traces without sacrificing generated same/different
verdict accuracy.
SpeakerLLM provides a controlled starting point for speaker-specialized
audio-LLMs that expose speaker evidence in language rather than only as backend
scores.


\vspace{-3mm}
\section{Limitations and Future Work}
\vspace{-2mm}
\label{sec:limitations}

This work studies controlled LLM-interface speaker evidence modeling.
Our faithfulness analysis is grounded in the supervised reasoning schema rather
than causal internal mechanisms; future work should add counterfactual,
intervention-based, and human-grounded evaluations, and connect the language
interface to calibrated score-based SV backends when thresholded operation is
required.
Scaling to larger in-the-wild corpora, broader languages and accents, and real
noisy or far-field recordings is another natural extension.
More broadly, voice-only agents require consent-aware interfaces for user
enrollment, authorization, personalization, and speaker-aware user
understanding.
Because the framework uses biometric speaker identity and profile attributes,
deployment should include consent-aware data use, privacy-preserving inference,
and fairness evaluation across demographic and acoustic conditions.

\bibliographystyle{unsrtnat}
\bibliography{references}

\newpage
\clearpage
\appendix

\newpage
\section*{Technical Appendices and Supplementary Material}

This supplementary material complements the main paper by providing
the following sections.
To support code reproducibility, we will release the source code
along with a README file.

\vspace{16pt}

\renewcommand{\contentsname}{}

\noindent
\begin{tabularx}{\linewidth}{@{} X @{\quad} r @{}}

\textbf{A} \quad \textbf{Attribute and Metadata Details}
  & \pageref{app:metadata} \\[4pt]
\quad A.1 \quad Detailed Attribute Taxonomy and Binning
  & \pageref{app:attribute-binning} \\[2pt]
\quad A.2 \quad Environment Label Computation
  & \pageref{app:env-label-computation} \\[2pt]
\quad A.3 \quad Acoustic Feature Extraction
  & \pageref{app:acoustic-feature-extraction} \\[2pt]
\quad A.4 \quad Metadata Coverage
  & \pageref{app:metadata-coverage} \\[10pt]

\textbf{B} \quad \textbf{Supervision Design and Target Construction}
  & \pageref{app:details} \\[4pt]
\quad B.1 \quad Stage-1 Target Templates
  & \pageref{app:stage1-templates} \\[2pt]
\quad B.2 \quad Stage-2 Target Templates
  & \pageref{app:stage2-templates} \\[2pt]
\quad B.3 \quad Profile-Support Computation
  & \pageref{app:profile-support} \\[2pt]
\quad B.4 \quad Decision Composition Rules
  & \pageref{app:decision-rulebook} \\[2pt]
\quad B.5 \quad Worked Examples
  & \pageref{app:worked-examples} \\[10pt]

\textbf{C} \quad \textbf{Details of Experimental Settings}
  & \pageref{app:exp-settings} \\[4pt]
\quad C.1 \quad Dataset Statistics
  & \pageref{app:dataset-stats} \\[2pt]
\quad C.2 \quad Model and Speaker-Tokenizer Hyperparameters
  & \pageref{app:model-hparams} \\[2pt]
\quad C.3 \quad Optimization Schedule
  & \pageref{app:optimization} \\[2pt]
\quad C.4 \quad Augmentation Protocol
  & \pageref{app:augmentation} \\[2pt]
\quad C.5 \quad Evaluation Protocol
  & \pageref{app:eval-protocol} \\[2pt]
\quad C.6 \quad Baseline Prompting
  & \pageref{app:baseline-prompts} \\[10pt]

\textbf{D} \quad \textbf{Detailed Results}
  & \pageref{app:detailed-results} \\[4pt]
\quad D.1 \quad Speaker-Token Count Ablation
  & \pageref{app:tokenizer-token-count} \\[2pt]
\quad D.2 \quad Per-Attribute Analysis of Speaker-Tokenizer Warm-Up
  & \pageref{app:tokenizer-warmup-ablation} \\[2pt]
\quad D.3 \quad Representative SV-R Cases
  & \pageref{app:qual-examples} \\

\end{tabularx}

\newpage

\section{Attribute and Metadata Details}
\label{app:metadata}

This appendix specifies the attribute definitions used to construct
SpeakerLLM supervision. It provides the binning boundaries, acoustic
feature extraction procedure, and per-corpus metadata coverage underlying
the attribute taxonomy summarized in the main paper.

\subsection{Detailed Attribute Taxonomy and Binning}
\label{app:attribute-binning}

Table~\ref{tab:app-attribute-binning} lists the class definitions and
binning rules used to convert metadata, acoustic measurements, and simulated
recording conditions into discrete supervision labels.

\begin{table*}[t]
\centering
\small
\renewcommand{\arraystretch}{1.2}
\caption{%
Detailed attribute taxonomy and binning rules used for supervision label
construction.
Pitch is discretized using gender-conditioned F0 percentile cutoffs, while
timbral brightness uses gender-shared spectral-centroid percentile cutoffs.
Noise and reverberation labels are generated by controlled simulation using
SNR and pre-computed RT60, respectively.
}
\label{tab:app-attribute-binning}
\setlength{\tabcolsep}{5pt}
\begin{tabularx}{\textwidth}{@{}l l X@{}}
\toprule
\textbf{Attribute} & \textbf{Class} & \textbf{Boundary / rule} \\
\midrule

\multicolumn{3}{@{}l}{\textit{Speaker profile}} \\
\addlinespace[2pt]

Gender
& male / female
& From public metadata (VoxCeleb Enrichment, LibriTTS-P). \\
\addlinespace[3pt]

Age
& 10 ordinal bins
& 1--7, 8--12, 13--17, 18--25, 26--35, 36--45, 46--55, 56--65, 66--75, 76+.
  Derived from VoxCeleb Enrichment age estimates. \\
\addlinespace[3pt]

Region
& 8 classes
& Nationality $\rightarrow$ regional/linguistic background mapping:
  North American, European, British/Irish, Latin/Hispanic,
  Oceanian, East/Southeast Asian, Middle Eastern/North African, African. \\
\addlinespace[3pt]

Pitch
& 5 gender-conditioned bins
& Per-utterance mean F0 is ranked within each gender's pooled distribution.
  Percentile cutoffs at 10\%, 30\%, 70\%, 90\% yield
  \texttt{very low}, \texttt{low}, \texttt{normal}, \texttt{high},
  \texttt{very high}. \\
\addlinespace[3pt]

Timbral brightness
& 5 bins (gender-shared)
& Per-utterance mean spectral centroid is ranked within the pooled distribution.
  Same percentile cutoffs (10/30/70/90\%) yield
  \texttt{muted}, \texttt{mellow}, \texttt{neutral}, \texttt{bright},
  \texttt{brilliant}. \\

\midrule
\multicolumn{3}{@{}l}{\textit{Environment (online simulation)}} \\
\addlinespace[2pt]

Noise
& 5 SNR-controlled classes
& \texttt{clean} ($\geq 20$ dB),
  \texttt{mild} ($[10,20)$ dB),
  \texttt{moderate} ($[5,10)$ dB),
  \texttt{severe} ($[0,5)$ dB),
  \texttt{extreme} ($<0$ dB). \\
\addlinespace[3pt]

Reverberation
& 5 RT60-controlled classes
& \texttt{minimal} ($\leq 0.3$ s),
  \texttt{slight} ($(0.3,0.6]$ s),
  \texttt{moderate} ($(0.6,1.0]$ s),
  \texttt{heavy} ($(1.0,1.5]$ s),
  \texttt{extreme} ($>1.5$ s). \\

\bottomrule
\end{tabularx}
\end{table*}

\subsection{Environment Label Computation}
\label{app:env-label-computation}

Noise labels are controlled by the signal-to-noise ratio (SNR) between the
clean speech waveform $s$ and the scaled non-speech noise waveform $\alpha n$:
\[
\mathrm{SNR}_{\mathrm{dB}}
=
10 \log_{10}
\frac{\sum_t s_t^2}{\sum_t (\alpha n_t)^2}.
\]
The scale $\alpha$ is chosen to match the target SNR interval.
For reverberation, we estimate broadband RT60 for each SLR28 room impulse
response using Schroeder backward integration~\cite{schroeder1965reverberation}.
We fit a T30-style line to the $-5$ to $-35$\,dB region of the energy decay
curve and extrapolate it to 60\,dB decay.
The resulting RIR-level RT60 estimate is used for reverberation bucketing
(Table~\ref{tab:app-attribute-binning}); it is not intended as a full
room-acoustic measurement protocol.

\subsection{Acoustic Feature Extraction}
\label{app:acoustic-feature-extraction}

Pitch and timbral brightness are extracted from audio for both VoxCeleb~\cite{nagrani17_interspeech}
and LibriTTS-R~\cite{koizumi23_interspeech} utterances.
Pitch is derived from the per-utterance mean fundamental frequency (F0),
estimated using \texttt{librosa.pyin}.
Timbral brightness is derived from the per-utterance mean spectral centroid
computed over STFT frames.
Because F0 distributions differ substantially between male and female speakers,
pitch is discretized using gender-conditioned percentile cutoffs
(Table~\ref{tab:app-attribute-binning}), so that a given pitch label reflects
relative position within the speaker's gender group rather than an absolute
frequency range.
These acoustic descriptors are used only for training instances that require
pitch- or brightness-conditioned labels. Approximately 11,000 utterances
($<1\%$ of the combined corpus) failed F0 extraction and are therefore
excluded from pitch- and brightness-conditioned instances, while remaining
available for tasks that do not require these descriptors.

\subsection{Metadata Coverage}
\label{app:metadata-coverage}

Table~\ref{tab:app-metadata-coverage} reports the fraction of utterances
for which each speaker-profile attribute is available, broken down by source
corpus. Because metadata availability is partial, attribute-specific
training instances are generated only when the corresponding field exists.
Missing attributes are not imputed.

\begin{table}[ht]
\centering
\small
\renewcommand{\arraystretch}{1.15}
\caption{%
Per-corpus metadata coverage used for profile-label construction.
Pitch and timbral brightness are audio-derived descriptors; their coverage is
reported after excluding utterances for which acoustic feature extraction
fails.
}
\label{tab:app-metadata-coverage}
\setlength{\tabcolsep}{5pt}
\begin{tabular}{@{}l c c@{}}
\toprule
\textbf{Attribute} & \textbf{VoxCeleb} & \textbf{LibriTTS-R} \\
\midrule
Gender              & 99.8\% & 99.7\% \\
Age                 & 86.6\% & 0\%    \\
Region              & 99.2\% & 0\%    \\
Pitch               & 99.6\% & 98.4\% \\
Timbral brightness  & 99.6\% & 98.4\% \\
\bottomrule
\end{tabular}
\end{table}

\section{Supervision Design and Target Construction}
\label{app:details}

This appendix documents how SpeakerLLM training targets are constructed.
It covers three components:
(i)~the fixed target templates used in the Speaker Understanding stage and
the Verification-Reasoning Tuning stage
(Tables~\ref{tab:stage1-templates}--\ref{tab:stage2-templates}),
(ii)~the penalty-based profile-support computation used by the
verification reasoning target (Table~\ref{tab:profile-support}), and
(iii)~the decision composition rules that map environment severity,
profile-support level, and the ground-truth same/different label into the
final \textsc{decision} block
(Table~\ref{tab:decision-matrix}).

\subsection{Stage-1 Target Templates}
\label{app:stage1-templates}

Stage~1 supports Speaker Understanding and uses the same scenario inventory
across two target styles.
During speaker-tokenizer warm-up, short-form targets are used to stabilize
cue decoding and align audio-derived speaker representations with the language
model input space.
During sentence adaptation, the same scenarios are expressed as sentence-form
targets.
Table~\ref{tab:stage1-templates} lists the template for each scenario.
Slots enclosed in braces (e.g., \texttt{\{gender\}}) are filled from the
available per-utterance labels at training time.

\begin{table}[t!]
\centering
\small
\renewcommand{\arraystretch}{1.25}
\caption{%
Stage~1 target templates for Speaker Understanding.
\textbf{Short form} targets are used during speaker-tokenizer warm-up, and
\textbf{sentence form} targets are used during sentence adaptation.
The \emph{full speaker profile} and \emph{joint acoustic profile} targets are
constructed by concatenating the available atomic sentences from the
corresponding attribute rows.
}
\label{tab:stage1-templates}
\setlength{\tabcolsep}{4pt}
\begin{tabularx}{\linewidth}{@{}l l X X@{}}
\toprule
\textbf{Group} & \textbf{Scenario} & \textbf{Short form} & \textbf{Sentence form} \\
\midrule

\multicolumn{4}{@{}l}{\textit{Speaker profile (single-utterance)}} \\
\addlinespace[2pt]

& Gender
& \texttt{\{gender\}}
& \texttt{The speaker's gender is inferred to be \{gender\}.} \\

& Age
& \texttt{\{age\_phrase\}}
& \texttt{The speaker's age is \{age\_phrase\}.} \\

& Region
& \texttt{\{region\}}
& \texttt{The speaker's regional background is \{region\}.} \\

& Voice characteristic
& \texttt{\{brightness\} voice and \{pitch\} \{gender\}-range pitch}
& \texttt{The speaker has a \{brightness\} voice and \{pitch\} \{gender\}-range pitch.} \\

& Full speaker profile
& comma-separated atomic phrases
& Concatenation of available atomic sentences:\newline
  \texttt{The speaker is \{gender\}. The speaker is likely \{age\_phrase\}. The speaker has a \{region\} regional background. The speaker has a \{brightness\} voice and \{pitch\} \{gender\}-range pitch.} \\

\addlinespace[4pt]
\multicolumn{4}{@{}l}{\textit{Environment (single-utterance)}} \\
\addlinespace[2pt]

& Noise class
& \texttt{\{noise\_class\}}
& \texttt{The recording has \{noise\_class\} noise.} \\

& Reverberation class
& \texttt{\{reverb\_class\}}
& \texttt{The recording has \{reverb\_class\} reverberation.} \\

& Joint acoustic profile
& \texttt{\{noise\} noise and \{reverb\} reverberation}
& \texttt{The recording has \{noise\_class\} noise and \{reverb\_class\} reverberation.} \\

\addlinespace[4pt]
\multicolumn{4}{@{}l}{\textit{Utterance-pair tasks}} \\
\addlinespace[2pt]

& standard SV
& \texttt{same} / \texttt{different}
& \texttt{These recordings are from the same speaker.}\newline
  \texttt{These recordings are from different speakers.} \\

& Noise comparison
& \texttt{speech1} / \texttt{speech2} / \texttt{similar}
& \texttt{Speech 1 is noisier.} / \texttt{Speech 2 is noisier.}\newline
  \texttt{Both recordings have similar noise levels.} \\

& Reverb comparison
& \texttt{speech1} / \texttt{speech2} / \texttt{similar}
& \texttt{Speech 1 is more reverberant.} / \texttt{Speech 2 is more reverberant.}\newline
  \texttt{Both recordings have similar reverberation levels.} \\

\bottomrule
\end{tabularx}
\end{table}

\subsection{Stage-2 Target Templates}
\label{app:stage2-templates}

Stage~2 supports Verification-Reasoning Tuning and introduces two task
families: attribute compatibility QA and verification reasoning targets.
All Stage~2 targets are sentence-form from the outset.
Table~\ref{tab:stage2-templates} lists the templates.
The \textsc{profile\_compatibility} block of the verification reasoning
target reuses the same atomic clause builders as the compatibility QA tasks
(e.g., gender, age, region, and voice clauses), but omits the holistic
summary sentence.
The overall profile summary is instead composed inside the
\textsc{decision} block using the profile-support level
(see \S\ref{app:decision-rulebook}).

Full speaker-profile and SV-R targets require the complete profile-attribute
set for the involved utterance(s).
Because age and region annotations are available only for VoxCeleb, these
targets are generated on VoxCeleb utterances or pairs.
The \textsc{environment\_status} block in SV-R is populated by applying the
same online noise/reverberation simulation used for environment supervision to
the corresponding VoxCeleb utterances.

\begin{table}[t!]
\centering
\small
\renewcommand{\arraystretch}{1.25}
\caption{%
Stage~2 target templates for Verification-Reasoning Tuning.
Compatibility QA targets compare one profile axis at a time, while the
holistic variant concatenates the available profile-comparison clauses and
adds an overall summary.
The verification reasoning target follows a three-block format:
\textsc{environment\_status}, \textsc{profile\_compatibility}, and
\textsc{decision}; the \textsc{decision} block is constructed according to
the composition rules in Table~\ref{tab:decision-matrix}.
}
\label{tab:stage2-templates}
\setlength{\tabcolsep}{4pt}
\begin{tabularx}{\linewidth}{@{}l l X@{}}
\toprule
\textbf{Family} & \textbf{Subtype} & \textbf{Target template} \\
\midrule

\multicolumn{3}{@{}l}{\textit{Attribute compatibility QA (utterance-pair comparison)}} \\
\addlinespace[2pt]

& Gender
& \texttt{Gender is similar.} ~~or~~ \texttt{Gender is different.} \\

& Age
& \texttt{The age ranges are similar.}\newline
  \texttt{The age ranges are slightly different.}\newline
  \texttt{The age ranges are very different.} \\

& Region
& \texttt{Linguistic background is similar.} or \texttt{Linguistic} \texttt{background} \texttt{is} \texttt{different.} \\

& Voice
& \texttt{Pitch is \{similar $|$ somewhat different $|$ very different\}.}\newline
  \texttt{Timbral brightness is \{similar $|$ somewhat different $|$ very different\}.}\newline
  \texttt{Therefore, the vocal characteristics are \{similar $|$ somewhat different $|$ very different\}.} \\

& Holistic
& Concatenation of all available atomic clauses above, followed by:\newline
  \texttt{Therefore, the overall speaker profile is \{similar $|$ somewhat different $|$ very different\}.} \\

\addlinespace[6pt]
\multicolumn{3}{@{}l}{\textit{Verification reasoning target (three-block format)}} \\
\addlinespace[2pt]

& \textsc{environment\_status}
& \texttt{The first recording contains \{noise$_1$\} noise and \{reverb$_1$\} reverberation.}\newline
  \texttt{The second recording contains \{noise$_2$\} noise and \{reverb$_2$\} reverberation.} \\
\addlinespace[2pt]

& \textsc{profile\_compatibility}
& Same atomic clauses as compatibility QA, but \emph{without} the
  overall summary sentence. \\
\addlinespace[2pt]

& \textsc{decision}
& Composed from environment severity $\times$ profile-support level $\times$ ground-truth same/different label.\newline
  See Table~\ref{tab:decision-matrix} for the full composition rules. \\
\addlinespace[2pt]

& \textbf{Final target}
& \texttt{ENVIRONMENT\_STATUS:}\newline
  \texttt{\{environment\_status text\}}\newline\newline
  \texttt{PROFILE\_COMPATIBILITY:}\newline
  \texttt{\{profile\_compatibility text\}}\newline\newline
  \texttt{DECISION:}\newline
  \texttt{\{decision text\}} \\

\bottomrule
\end{tabularx}
\end{table}

\subsection{Profile-Support Computation}
\label{app:profile-support}

The verification reasoning target uses an internal
\emph{profile-support level} with respect to a same-speaker interpretation:
\{\textsc{supportive}, \textsc{mixed}, \textsc{conflicting}\}.
This level summarizes whether the available speaker-profile attributes support,
ambiguously mix with, or conflict with a same-speaker interpretation before
the final verification verdict is composed.
It is computed by summing attribute-level penalties over attributes that are
available for both utterances
(Table~\ref{tab:profile-support}).
The profile-support level is not itself the final verification label.
Instead, it provides an intermediate profile-evidence state used by the
decision composition rules in \S\ref{app:decision-rulebook}.
This separation allows the target to include reversal cases, where surface
profile evidence and the ground-truth verification label point in different
directions.

Gender receives the largest single penalty ($+4$), reflecting that a gender
mismatch often co-occurs with divergence in downstream vocal cues.
Region receives a deliberately smaller penalty ($+1$) to avoid
over-penalizing bilingual, multilingual, or cross-regional speakers.
Pitch and timbral brightness are treated as a grouped voice factor:
the voice penalty is the \emph{maximum} of the two sub-attribute penalties,
with an additional $+1$ bonus when both sub-attributes are non-compatible.
This captures the intuition that a joint mismatch in vocal range and timbral
color is stronger evidence than either cue alone.

For ordered attributes (age, pitch, timbral brightness), compatibility is
determined by bin distance: identical bins are \emph{compatible}
($p=0$), adjacent bins are \emph{partial} ($p=1$), and bins separated by
two or more steps are \emph{conflicting} ($p=2$).
For pitch, the ordered scale follows gender-conditioned F0 percentile bins
(Appendix~\ref{app:attribute-binning}); for timbral brightness, the order is
\texttt{muted} $<$ \texttt{mellow} $<$ \texttt{neutral} $<$ \texttt{bright}
$<$ \texttt{brilliant}.
Gender and region are binary: a mismatch incurs the penalty listed in
Table~\ref{tab:profile-support}; a match contributes zero.

\begin{table}[t]
\centering
\small
\renewcommand{\arraystretch}{1.15}
\caption{%
Profile-support scoring used by the verification reasoning target.
Penalties are summed over attributes that are comparable for both utterances;
missing attributes contribute no penalty.
The grouped voice factor combines pitch and timbral brightness using a
max-based penalty with an additional bonus for joint mismatch.
}
\label{tab:profile-support}
\setlength{\tabcolsep}{5pt}
\begin{tabular}{@{}l l c@{}}
\toprule
\textbf{Attribute} & \textbf{Condition} & \textbf{Penalty} \\
\midrule
Gender     & different           & $+4$ \\
\addlinespace[2pt]
Age        & adjacent bin        & $+1$ \\
           & gap $\ge 2$ bins    & $+2$ \\
\addlinespace[2pt]
Region     & different           & $+1$ \\
\addlinespace[2pt]
Voice (grouped)
           & one sub-attr non-compatible
           & $\max(p_\text{pitch}, p_\text{bright})$ \\
           & both sub-attrs non-compatible
           & $\max(p_\text{pitch}, p_\text{bright}) + 1$ \\
\addlinespace[1pt]
\multicolumn{3}{@{}l}{\quad\small\textit{where each sub-attr penalty $p \in \{0\text{ (compatible)},\; 1\text{ (partial)},\; 2\text{ (conflicting)}\}$}} \\
\midrule
\multicolumn{2}{@{}l}{\textsc{supportive}}  & total $\le 1$ \\
\multicolumn{2}{@{}l}{\textsc{mixed}}       & total $= 2$--$3$, or no fields \\
\multicolumn{2}{@{}l}{\textsc{conflicting}} & total $\ge 4$ \\
\bottomrule
\end{tabular}
\end{table}

\subsection{Decision Composition Rules}
\label{app:decision-rulebook}

The \textsc{decision} block is constructed from four phrases selected by
three input variables. These rules are used to construct supervision targets;
they are not provided to the model at inference time.

\begin{enumerate}[leftmargin=1.8em, itemsep=2pt]
\item \textbf{Environment clause} --- selected by pair severity
      (Table~\ref{tab:env-clause}).
\item \textbf{Profile summary} --- selected by profile-support level.
\item \textbf{Connector} --- selected by the alignment between
      profile-support level and the ground-truth same/different label.
\item \textbf{Verification verdict} --- selected by the same alignment.
\end{enumerate}

\noindent
The full \textsc{decision} is their concatenation:
\[
\textsc{decision}
= \underbrace{\text{env clause}}_{\text{pair severity}}
\;+\; \underbrace{\text{profile summary}
  \;+\; \text{connector}
  \;+\; \text{verification verdict}}_{\text{profile support} \times \text{GT label}}
\]
This construction separates profile-level evidence from the final
same/different verdict, allowing the target to express both aligned cases and
reversal cases in a linguistically coherent form.

\paragraph{Environment clause.}
The environment clause is selected from a three-way pair-severity level
(Table~\ref{tab:env-clause}). It describes how much the recording conditions
may degrade speaker-relevant cues and is independent of the profile-support
level and ground-truth verification label.

Pair severity is derived deterministically from the per-utterance environment
labels.
Each noise class and reverberation class is assigned an ordinal degradation
rank from~0 (clean/minimal) to~4 (extreme).
The degradation rank of a single recording is the maximum of its noise and
reverberation ranks.
The pair-severity rank is then the maximum degradation rank over the two
recordings.
Pair severity is defined as \emph{Low} if the pair rank is~0--1,
\emph{Moderate} if~2, and \emph{Extreme} if~3--4.

\begin{table}[ht]
\centering
\small
\renewcommand{\arraystretch}{1.15}
\caption{%
Environment clause selection for the first sentence of the
\textsc{decision} block.
Pair severity determines how strongly recording-condition mismatch or
degradation is described as affecting speaker-relevant cues.
}
\label{tab:env-clause}
\begin{tabular}{@{}l p{10cm}@{}}
\toprule
\textbf{Pair severity} & \textbf{Environment clause} \\
\midrule
Low
& Environmental mismatch or degradation is limited, so the speaker-relevant
  vocal cues remain clear. \\
Moderate
& Environmental mismatch or degradation is present, so the speaker-relevant
  vocal cues are partially degraded. \\
Extreme
& Strong environmental mismatch or severe degradation substantially
  weakens the speaker-relevant vocal cues. \\
\bottomrule
\end{tabular}
\end{table}

\paragraph{Profile summary, connector, and verification verdict.}
The remaining three phrases are jointly determined by profile-support level
$\times$ ground-truth same/different label, yielding six cases
(Table~\ref{tab:decision-matrix}).
The connector indicates whether the profile-level evidence and the final
verification verdict point in the same direction (\emph{Likewise}) or diverge
(\emph{However}). We mark reversal cases ($\star$), where profile evidence and
the ground-truth label disagree. These cases are included to prevent a shortcut
in which surface profile similarity mechanically determines the verification
verdict.

\begin{table}[t!]
\centering
\small
\renewcommand{\arraystretch}{1.3}
\caption{%
Decision composition matrix.
Each cell specifies the profile-summary clause, connector, and
verification-verdict clause for one combination of profile-support level and
ground-truth same/different label.
$\star$ marks reversal cases, where profile-level evidence and the final
verification verdict disagree.
}
\label{tab:decision-matrix}
\setlength{\tabcolsep}{5pt}
\begin{tabularx}{\linewidth}{@{}l X X@{}}
\toprule
& \textbf{GT = same speaker} & \textbf{GT = different speakers} \\
\midrule

\textbf{Supportive}
& \textit{Across the speaker profile, many attributes are similar.}
  \textbf{Likewise,}
  \textit{the latent speaker-identity cues also show strong similarity.
  Taken together, the recordings are determined to be from the same speaker.}
  \hfill\mbox{[aligned]}
& $\star$
  \textit{Across the speaker profile, many attributes are similar.}
  \textbf{However,}
  \textit{the latent speaker-identity cues show stronger separation.
  Taken together, the recordings are determined to be from different speakers.}
  \hfill\mbox{[reversal]}
\\
\addlinespace[4pt]

\textbf{Mixed}
& \textit{Across the speaker profile, some attributes are similar, while
  others differ.}
  \textit{The latent speaker-identity cues show stronger similarity.
  Taken together, the recordings are determined to be from the same speaker.}
  \hfill\mbox{[neutral]}
& $\star$
  \textit{Across the speaker profile, some attributes are similar, while
  others differ.}
  \textbf{However,}
  \textit{the latent speaker-identity cues show stronger separation.
  Taken together, the recordings are determined to be from different speakers.}
  \hfill\mbox{[reversal]}
\\
\addlinespace[4pt]

\textbf{Conflicting}
& $\star$
  \textit{Across the speaker profile, several attributes differ.}
  \textbf{However,}
  \textit{the latent speaker-identity cues show stronger similarity.
  Taken together, the recordings are determined to be from the same speaker.}
  \hfill\mbox{[reversal]}
& \textit{Across the speaker profile, several attributes differ.}
  \textbf{Likewise,}
  \textit{the latent speaker-identity cues also show clear differences.
  Taken together, the recordings are determined to be from different speakers.}
  \hfill\mbox{[aligned]}
\\

\bottomrule
\end{tabularx}
\end{table}

\paragraph{Connector logic.}
\emph{Likewise} is used when profile-level evidence and the ground-truth label
are aligned (\textsc{supportive}$+$same or
\textsc{conflicting}$+$different), indicating that latent speaker-identity
cues confirm the surface-level profile tendency.
\emph{However} is used in reversal cases and in the
\textsc{mixed}$+$different case, indicating that the final verdict departs
from what the profile evidence alone would suggest.
For \textsc{mixed}$+$same, no connector is inserted because the profile
evidence is already ambiguous and does not require contrastive framing.

\FloatBarrier

\subsection{Worked Examples}
\label{app:worked-examples}

Below are complete target examples produced by the verification reasoning
target construction policy. They illustrate an aligned supportive case, a
reversal case, and an aligned conflicting case. These examples are target
templates used for supervision, not model-generated outputs.

\vspace{0.6em}
\noindent
\fbox{\parbox{0.97\linewidth}{%
\textbf{Example 1 --- Aligned: supportive profile, same speaker}\\[0.3em]
\textsc{environment\_status}:\\
The first recording contains no background noise and minimal reverberation.
The second recording contains mild noise and slight reverberation.\\[0.4em]
\textsc{profile\_compatibility}:\\
Gender is similar. The age ranges are similar.
Linguistic background is similar.
Pitch is similar. Timbral brightness is similar.\\[0.4em]
\textsc{decision}:\\
Environmental mismatch or degradation is limited, so the speaker-relevant
vocal cues remain clear.
Across the speaker profile, many attributes are similar.
Likewise, the latent speaker-identity cues also show strong similarity.
Taken together, the recordings are determined to be from the same speaker.
}}

\vspace{0.8em}
\noindent
\fbox{\parbox{0.97\linewidth}{%
\textbf{Example 2 --- Reversal: supportive profile, different speakers
($\star$)}\\[0.3em]
\textsc{environment\_status}:\\
The first recording contains moderate noise and slight reverberation.
The second recording contains mild noise and moderate reverberation.\\[0.4em]
\textsc{profile\_compatibility}:\\
Gender is similar. The age ranges are similar.
Linguistic background is similar.
Pitch is somewhat different. Timbral brightness is similar.\\[0.4em]
\textsc{decision}:\\
Environmental mismatch or degradation is present, so the speaker-relevant
vocal cues are partially degraded.
Across the speaker profile, many attributes are similar.
However, the latent speaker-identity cues show stronger separation.
Taken together, the recordings are determined to be from different speakers.
}}

\vspace{0.8em}
\noindent
\fbox{\parbox{0.97\linewidth}{%
\textbf{Example 3 --- Aligned: conflicting profile, different speakers}\\[0.3em]
\textsc{environment\_status}:\\
The first recording contains severe noise and heavy reverberation.
The second recording contains severe noise and heavy reverberation.\\[0.4em]
\textsc{profile\_compatibility}:\\
Gender is similar. The age ranges are very different.
Linguistic background is similar.
Pitch is very different. Timbral brightness is somewhat different.\\[0.4em]
\textsc{decision}:\\
Strong environmental mismatch or severe degradation substantially weakens
the speaker-relevant vocal cues.
Across the speaker profile, several attributes differ.
Likewise, the latent speaker-identity cues also show clear differences.
Taken together, the recordings are determined to be from different speakers.
}}

\section{Details of Experimental Settings}
\label{app:exp-settings}

This appendix provides reproducibility details corresponding to
Section~\ref{sec:exp-settings}. It includes dataset statistics,
model and speaker-tokenizer hyperparameters, optimization schedules,
augmentation protocol, evaluation protocol, and baseline prompting.
Attribute definitions and binning boundaries are documented separately in
Appendix~\ref{app:metadata}.

\subsection{Dataset Statistics}
\label{app:dataset-stats}

Table~\ref{tab:app-dataset-detail} summarizes the training data used in the
controlled base-scale setting adopted for all main comparisons and ablations.
Table~\ref{tab:app-eval-sets} lists the evaluation sets used for SV,
speaker-profile QA, and environment QA.

\begin{table}[H]
\centering
\small
\caption{%
Training data used in the controlled base-scale setting.
Hours are measured from waveform headers.
Speaker counts are corpus-level counts because VoxCeleb and LibriTTS-R use
different speaker namespaces.
}
\label{tab:app-dataset-detail}
\setlength{\tabcolsep}{5pt}
\begin{tabular}{@{}l r r r@{}}
\toprule
\textbf{Corpus / list} & \textbf{Utt.} & \textbf{Spk.} & \textbf{Hours} \\
\midrule
VoxCeleb1-dev           & 148,642   & 1,211 & 340.4 \\
LibriTTS-R clean-360h   & 116,462   & 904   & 190.4 \\
\midrule
\textbf{Total}          & 265,104   & 2,115 & 530.8 \\
\bottomrule
\end{tabular}
\end{table}

\begin{table}[H]
\centering
\small
\caption{%
Evaluation sets.
VoxCeleb1-O provides the SV trials; speaker-profile QA is evaluated on the
unique utterances appearing in those trials.
LibriTTS-R test-clean is used for environment QA under controlled noise and
reverberation labels.
}
\label{tab:app-eval-sets}
\setlength{\tabcolsep}{5pt}
\begin{tabular}{@{}l r r r l@{}}
\toprule
\textbf{Set} & \textbf{Samples} & \textbf{Spk.} & \textbf{Hours} & \textbf{Used for} \\
\midrule
VoxCeleb1-O           & 37,611 trials & 40 & 10.8 & SV + profile QA \\
LibriTTS-R test-clean & 4,837 utt.    & 39 & 8.5  & environment QA \\
\bottomrule
\end{tabular}
\end{table}

\subsection{Model and Speaker-Tokenizer Hyperparameters}
\label{app:model-hparams}

Tables~\ref{tab:app-model-hparams} and~\ref{tab:app-param-counts} report the
model configuration and trainable parameter counts used in the main
experiments.

\begin{table}[H]
\centering
\small
\caption{%
Model and speaker-tokenizer hyperparameters.
The speaker encoder (ReDimNet-B3~\cite{yakovlev24_interspeech}) is frozen, while the speaker-token adapter and LoRA
parameters are trained according to the phase schedule in
Table~\ref{tab:app-training-schedule}.
}
\label{tab:app-model-hparams}
\setlength{\tabcolsep}{6pt}
\begin{tabular}{@{}l l@{}}
\toprule
\textbf{Component} & \textbf{Setting} \\
\midrule
Speaker encoder          & ReDimNet-B3, frozen, pretrained on VoxCeleb2 \\
Speaker embedding dim    & 192 \\
Frame-level feature dim  & 1,152 \\
Language model           & Qwen2.5-1.5B-Instruct \\
LM hidden dim            & 1,536 \\
\addlinespace[3pt]
Embedding-level tokens (MLP prefix) & 16 \\
Sequence-level tokens (Q-Former)    & 32 \\
Total speaker tokens per utterance  & 48 \\
Q-Former hidden size                & 512 \\
Q-Former layers                     & 4 \\
Q-Former attention heads            & 8 \\
Q-Former intermediate size          & 2,048 \\
\addlinespace[3pt]
LoRA target modules     & \texttt{q\_proj}, \texttt{k\_proj}, \texttt{v\_proj}, \texttt{o\_proj} \\
LoRA rank / alpha / dropout & 16 / 32 / 0.05 \\
\bottomrule
\end{tabular}
\end{table}

\begin{table}[H]
\centering
\small
\caption{%
Trainable parameter counts.
The speaker encoder and base language-model weights remain frozen; trainable
parameters come from the speaker tokenizer and LoRA modules.
}
\label{tab:app-param-counts}
\setlength{\tabcolsep}{6pt}
\begin{tabular}{@{}l r@{}}
\toprule
\textbf{Module} & \textbf{Parameters} \\
\midrule
Speaker tokenizer        & 21.2M \\
LoRA                     & 5.5M \\
Speaker tokenizer + LoRA & 26.7M \\
\bottomrule
\end{tabular}
\end{table}

\subsection{Optimization Schedule}
\label{app:optimization}

Table~\ref{tab:app-training-schedule} reports the per-phase training schedule.
Table~\ref{tab:app-optim-common} lists optimization settings shared across
all phases.

\begin{table}[H]
\centering
\footnotesize
\caption{%
Training schedule across speaker-tokenizer warm-up, sentence adaptation,
and Verification-Reasoning Tuning.
The first phase trains only the speaker tokenizer; the later phases train the
speaker tokenizer together with LoRA parameters.
}
\label{tab:app-training-schedule}
\setlength{\tabcolsep}{3pt}
\begin{tabular}{@{}l l l r r r r r@{}}
\toprule
\textbf{Phase} & \textbf{Trainable} & \textbf{Targets}
& \textbf{Steps} & \textbf{Batch} & \textbf{Peak LR} & \textbf{Min LR} & \textbf{Warmup} \\
\midrule
Understanding
& tokenizer
& short-form
& 126,240
& 42
& $1.0{\times}10^{-4}$
& $5.0{\times}10^{-5}$
& 6,312 \\
Sent.\ adapt.
& tok.\ + LoRA
& sentence-form
& 82,840
& 32
& $4.0{\times}10^{-6}$
& $2.0{\times}10^{-6}$
& 4,142 \\
VR tuning
& tok.\ + LoRA
& verif.\ reasoning + compat.
& 220,920
& 24
& $6.0{\times}10^{-6}$
& $3.0{\times}10^{-6}$
& 11,046 \\
\bottomrule
\end{tabular}
\end{table}

\begin{table}[H]
\centering
\small
\caption{%
Optimization settings shared across all training phases.
}
\label{tab:app-optim-common}
\setlength{\tabcolsep}{6pt}
\begin{tabular}{@{}l l@{}}
\toprule
\textbf{Setting} & \textbf{Value} \\
\midrule
Optimizer       & AdamW \\
Weight decay    & $1.0{\times}10^{-2}$ \\
Precision       & bf16 mixed precision \\
Attention       & FlashAttention-2 \\
Scheduler       & cosine annealing with warmup \\
Hardware        & A6000 48\,GB $\times$ 4 \\
\bottomrule
\end{tabular}
\end{table}

\subsection{Augmentation Protocol}
\label{app:augmentation}

Noise and reverberation labels are generated online through acoustic
augmentation. The severity class definitions for these labels are provided in
Appendix~\ref{app:attribute-binning}.

For VoxCeleb utterances, noise and reverberation are each applied with
probability 0.5, yielding approximately equal proportions of noise-only,
reverberation-only, joint, and clean conditions.
For LibriTTS-R utterances, noise-only, reverberation-only, and joint
augmentation are each applied with probability 0.3, with the remaining
0.1 probability left as clean.
When both factors are applied, reverberation is convolved first, followed
by additive noise.

\subsection{Evaluation Protocol}
\label{app:eval-protocol}

Table~\ref{tab:app-eval-protocol} summarizes the evaluation protocol for each
task group.

\begin{table}[H]
\centering
\small
\caption{%
Evaluation protocol by task group.
All tasks are evaluated as generated-answer accuracy after deterministic
label parsing. SV denotes the standard same/different verdict mode, while
SV-R uses the final verdict parsed from the generated \textsc{decision} block.
}
\label{tab:app-eval-protocol}
\setlength{\tabcolsep}{6pt}
\begin{tabular}{@{}l l l@{}}
\toprule
\textbf{Task} & \textbf{Output} & \textbf{Metric} \\
\midrule
SV              & same/different verdict & accuracy \\
SV-R            & verdict in \textsc{decision} block & accuracy \\
Profile QA      & closed attribute label & accuracy \\
Environment QA  & closed condition label & accuracy \\
\bottomrule
\end{tabular}
\end{table}

Parsing failures are counted as incorrect.
At evaluation, when an utterance exceeds 15 seconds, we use a center crop
of 15 seconds; shorter utterances are used in full.
Metadata is used only for label construction and offline analysis, and is not
provided to models at inference time.

\subsection{Baseline Prompting}
\label{app:baseline-prompts}

Table~\ref{tab:app-baseline-prompts} summarizes the prompting protocol used
for each baseline model.

\begin{table}[H]
\centering
\footnotesize
\caption{%
Baseline prompting protocol.
General audio-LLMs are evaluated with our closed-option prompts, while
CoLMbo~\cite{baali2025colmbo} is evaluated with its native prompt for supported profiling attributes.
SA-TinyLLaMA~\cite{thebaud2026speaker} is listed for scope comparison because it reports score-based
EER rather than prompted generated-answer accuracy.
}
\label{tab:app-baseline-prompts}
\setlength{\tabcolsep}{4pt}
\begin{tabular}{@{}l l l@{}}
\toprule
\textbf{Baseline} & \textbf{Protocol} & \textbf{Notes} \\
\midrule
Qwen2.5-Omni-7B   & our closed-option prompts & class options provided \\
Qwen3.0-Omni-30B  & our closed-option prompts & class options provided \\
Audio Flamingo3     & our closed-option prompts & class options provided \\
CoLMbo             & native prompt where supported & gender, age, and region only \\
SA-TinyLLaMA       & not evaluated under our prompted protocol & reported score-based EER \\
\bottomrule
\end{tabular}
\end{table}

For dataset-specific labels such as pitch, timbral brightness, noise, and
reverberation, we include the class options and short class definitions in the
prompt (see Appendix~\ref{app:attribute-binning} for the class taxonomy).

\section{Detailed Results}
\label{app:detailed-results}

This appendix expands the analyses summarized in the main paper.
Section~\ref{app:tokenizer-token-count} controls the number of speaker tokens
to test whether the hierarchical tokenizer's gain can be explained by token
count alone.
Section~\ref{app:tokenizer-warmup-ablation} provides a per-attribute
breakdown of the speaker-tokenizer warm-up ablation.
Section~\ref{app:qual-examples} provides representative generated SV-R
traces.

\subsection{Speaker-Token Count Ablation}
\label{app:tokenizer-token-count}

Table~\ref{tab:speaker_tokenizer} in the main paper compares speaker-token
adapter architectures under their default token configurations.
Here, we additionally control the number of speaker tokens
for MLP-only and Q-Former-only adapters to test whether the hierarchical
tokenizer's gain is attributable to token count alone.
All variants are trained under Stage~1 speaker-tokenizer-only training with
the same frozen ReDimNet-B3~\cite{yakovlev24_interspeech} encoder and
Qwen2.5-1.5B~\cite{qwen25report} backbone.

\begin{table}[H]
\centering
\footnotesize
\caption{%
Speaker-token count ablation by adapter type.
MLP-only and Q-Former-only adapters are evaluated with 32 and 48 speaker
tokens.
The hierarchical speaker tokenizer uses 16 embedding-level tokens and
32 sequence-level tokens, for a total of 48 speaker tokens.
All values are accuracy (\%\,$\uparrow$) under Stage~1
speaker-tokenizer-only training.
}
\label{tab:app-token-ablation}
\setlength{\tabcolsep}{3pt}
\begin{tabular}{@{}l cc r c ccccc cc@{}}
\toprule
& \multicolumn{2}{c}{\textbf{Speaker Repr.}}
& &
& \multicolumn{5}{c}{\textbf{Speaker Profile}}
& \multicolumn{2}{c}{\textbf{Environment}} \\
\cmidrule(lr){2-3} \cmidrule(lr){6-10} \cmidrule(l){11-12}
\textbf{Adapter}
& Embed.
& Frames
& Tokens
& SV
& Gender
& Age
& Region
& Pitch
& Bright.
& Noise
& Reverb \\
\midrule
MLP
  & \checkmark &            & 32
  & 86.5 & 99.1 & 30.2 & 78.1 & 57.4 & 41.3 & 32.6 & 32.5 \\
MLP
  & \checkmark &            & 48
  & 93.7 & 98.9 & 40.5 & 80.7 & 52.3 & 37.6 & 30.3 & 32.2 \\
\addlinespace[3pt]
Q-Former
  &            & \checkmark & 32
  & 84.7 & 99.3 & 32.2 & 74.4 & 70.0 & 51.0 & 48.9 & 45.6 \\
Q-Former
  &            & \checkmark & 48
  & 90.7 & 99.4 & 38.3 & 82.8 & 71.7 & 51.0 & 47.1 & 48.0 \\
\midrule
Ours (MLP\,+\,Q-Former)
  & \checkmark & \checkmark & 16\,+\,32
  & \textbf{95.6} & \textbf{99.7} & 39.5 & 79.8 & \textbf{72.3} & \textbf{53.1} & 47.7 & \textbf{50.0} \\
\bottomrule
\end{tabular}
\end{table}

Increasing the number of speaker tokens from 32 to 48 improves SV accuracy
for both single-source adapters: MLP improves from $86.5$ to $93.7$, and
Q-Former improves from $84.7$ to $90.7$.
Thus, additional token capacity helps speaker-identity discrimination.
However, the attribute-level patterns show that capacity alone does not remove
the limitations of each representation source.
For MLP, increasing token count improves SV and identity-correlated attributes
such as age ($30.2 \rightarrow 40.5$) and region
($78.1 \rightarrow 80.7$), but pitch and timbral brightness degrade
($57.4 \rightarrow 52.3$ and $41.3 \rightarrow 37.6$).
For Q-Former, increasing token count improves SV and most profile attributes,
but the 48-token Q-Former still remains below the hierarchical tokenizer in
SV accuracy ($90.7$ vs.\ $95.6$) and is weaker on global attributes such as
age.
These results support the representation-granularity hypothesis: the
hierarchical tokenizer improves performance by combining embedding-level and
sequence-level evidence, rather than by token count alone.

\subsection{Per-Attribute Analysis of Speaker-Tokenizer Warm-Up}
\label{app:tokenizer-warmup-ablation}

Table~\ref{tab:app-tokenizer-warmup-ablation} expands the tokenizer warm-up
ablation summarized in Table~\ref{tab:curriculum_ablation}(a) of the main
paper.
It compares SpeakerLLM-Base against a variant that skips the initial
speaker-tokenizer-only warm-up and trains the speaker tokenizer together with
LoRA-adapted language-model parameters from the start.

\begin{table}[H]
\centering
\caption{%
Per-attribute breakdown of the speaker-tokenizer warm-up ablation.
SpeakerLLM-Base includes the initial speaker-tokenizer-only warm-up, whereas
the ablated variant trains the speaker tokenizer and LoRA-adapted language
model jointly from the start.
All values are accuracy (\%\,$\uparrow$).
}
\label{tab:app-tokenizer-warmup-ablation}
\setlength{\tabcolsep}{4pt}
\begin{tabular}{@{}l c ccccc cc@{}}
\toprule
& & \multicolumn{5}{c}{\textbf{Speaker Profile}}
& \multicolumn{2}{c}{\textbf{Environment}} \\
\cmidrule(lr){3-7} \cmidrule(l){8-9}
\textbf{Setting}
& SV
& Gender
& Age
& Region
& Pitch
& Bright.
& Noise
& Reverb \\
\midrule
w/o warm-up
& 91.20 & 98.98 & 34.34 & 78.40 & 65.60 & 43.96 & 38.05 & 42.86 \\
SpeakerLLM-Base
& \textbf{96.05} & \textbf{99.89} & \textbf{39.83} & \textbf{83.14} & \textbf{72.44} & \textbf{54.19} & \textbf{52.70} & \textbf{51.72} \\
\bottomrule
\end{tabular}
\end{table}

Speaker-tokenizer warm-up improves both speaker verification and attribute
understanding.
SV accuracy improves from $91.20$ to $96.05$, while non-trivial
speaker-profile attributes also improve: age
($34.34 \rightarrow 39.83$), region ($78.40 \rightarrow 83.14$), pitch
($65.60 \rightarrow 72.44$), and timbral brightness
($43.96 \rightarrow 54.19$).
The largest gains appear for fine-grained acoustic or recording-condition
attributes: noise ($38.05 \rightarrow 52.70$) and reverberation
($42.86 \rightarrow 51.72$).
These patterns indicate that warm-up is most beneficial for cues that require
stable alignment between audio-derived speaker representations and the
language-model input space.

Gender changes only slightly ($98.98 \rightarrow 99.89$) because it is
already near saturation.
In contrast, pitch, timbral brightness, noise, and reverberation are more
sensitive to whether the speaker-token interface is stabilized before LoRA
adaptation.
Thus, the warm-up phase functions as an alignment stage, rather than a minor
optimization detail.

\subsection{Representative Speaker Verification Reasoning Cases}
\label{app:qual-examples}

Table~\ref{tab:qual_examples} provides representative generated SV-R traces
for two difficult trial types.
The metadata and profile-support states are not provided to the model at
inference time; they are shown only as offline evidence for comparing the
generated SV-R output with the target evidence schema.

\begin{table}[t]
\centering
\footnotesize
\caption{
Representative cases where SV fails but SV-R is correct.
Metadata and profile-support states are for offline analysis only.
Dec.\ neg.\ = profile-deceptive negative;
Deg.\ pos.\ = degradation-aware positive.
}
\label{tab:qual_examples}
\setlength{\tabcolsep}{3pt}
\begin{tabularx}{\linewidth}{@{}p{1.6cm} c cc >{\hsize=0.7\hsize}X >{\hsize=1.3\hsize}X@{}}
\toprule
\textbf{Case}
& \textbf{GT}
& \textbf{SV}
& \textbf{SV-R}
& \textbf{Offline evidence}
& \textbf{SV-R decision excerpt} \\
\midrule

Dec.\ neg.
& diff.
& same
& diff.
& All profile attributes (gender, age, region, pitch, brightness) are compatible.
& ``Many attributes are similar. However, the latent speaker-identity cues show
stronger separation. \ldots{} determined to be from different speakers.'' \\

\addlinespace[4pt]

Deg.\ pos.
& same
& diff.
& same
& Gender, age, region compatible; pitch and brightness differ slightly;
moderate noise in first recording.
& ``Environmental degradation is present \ldots{} vocal cues partially degraded.
The latent speaker-identity cues show strong similarity. \ldots{} determined to
be from the same speaker.'' \\

\bottomrule
\end{tabularx}
\end{table}

In the \emph{profile-deceptive negative} case, the offline profile evidence is
highly supportive: gender, age, region, pitch, and timbral brightness are all
compatible, yet the ground-truth label is different.
Standard SV predicts ``same'', whereas SV-R first exposes the similar
profile-level evidence and then predicts ``different'' based on stronger
identity-level separation.

In the \emph{degradation-aware positive} case, the two utterances are from the
same speaker, but one recording contains moderate noise and the voice
characteristics partially differ.
Standard SV predicts ``different'', whereas SV-R describes the degradation and
then preserves the same-speaker decision based on identity-level speaker
evidence.

\end{document}